\documentclass{aastex}

\usepackage{amsmath}
\usepackage[TS1,T1]{fontenc}\usepackage{textcomp}
\usepackage{graphicx}

\usepackage{emulateapj5}

\usepackage{timesfonts}
\renewcommand{\v}{V}

\makeatletter
\newenvironment{inlinetable}{%
\def\@captype{table}%
\noindent\begin{minipage}{0.999\linewidth}\begin{center}\footnotesize}
{\end{center}\end{minipage}\smallskip}

\newenvironment{inlinefigure}{%
\def\@captype{figure}%
\noindent\begin{minipage}{0.999\linewidth}\begin{center}}
{\end{center}\end{minipage}\smallskip}
\makeatother

\newcommand{\keV}{ke\kern-0.05emV}
\newcommand{\HF}{{}_2\!F\!{}_1}
\newcommand{\degree}{\degr}

\begin{document}

\title{A moving cold front in the intergalactic medium of A3667.}

\author{A.\ Vikhlinin\altaffilmark{1}, M.\ Markevitch, S.\ S.\ Murray}
\affil{Harvard-Smithsonian Center for Astrophysics, 60 Garden St.,
Cambridge, MA 02138;\\ avikhlinin@cfa.harvard.edu}

\altaffiltext{1}{Also Space Research Institute, Moscow, Russia}

\shorttitle{COLD FRONT IN A3667}
\shortauthors{VIKHLININ, MARKEVITCH, \& MURRAY}

\begin{abstract}
  
  We present results from a \emph{Chandra} observation of the central region
  of the galaxy cluster A3667, with emphasis on the prominent sharp X-ray
  brightness edge spanning 0.5~Mpc near the cluster core. Our temperature
  map shows large-scale nonuniformities characteristic of the ongoing
  merger, in agreement with earlier \textsl{ASCA} results. The brightness
  edge turns out to be a boundary of a large cool gas cloud moving through
  the hot ambient gas, very similar to the ``cold fronts'' discovered by
  \emph{Chandra} in A2142.  The higher quality of the A3667 data allows the
  direct determination of the cloud velocity. At the leading edge of the
  cloud, the gas density abruptly increases by a factor of $3.9\pm0.8$,
  while the temperature decreases by a factor of $1.9\pm0.2$ (from 7.7~\keV\ 
  to 4.1~\keV). The ratio of the gas pressures inside and outside the front
  shows that the cloud moves through the ambient gas at near-sonic velocity,
  $M=1\pm0.2$ or $\v=1400\pm300$~km~s$^{-1}$. In front of the cloud, we
  observe the compression of the ambient gas with an amplitude expected for
  such a velocity. A smaller surface brightness discontinuity is observed
  further ahead, $\sim350$~kpc in front of the cloud. We suggest that it
  corresponds to a weak bow shock, implying that the cloud velocity may be
  slightly supersonic. Given all the evidence, the cold front appears to
  delineate the remnant of a cool subcluster that recently has merged with
  A3667. The cold front is remarkably sharp.  The upper limit on its width,
  $3.5''$ or 5~kpc, is several times smaller than the Coulomb mean free
  path. This is a direct observation of suppression of the transport
  processes in the intergalactic medium, most likely by magnetic fields.
  
\end{abstract}


\section{Introduction}

%

\begin{figure*}[htb]
  \centerline{
    \includegraphics[width=0.85\linewidth]{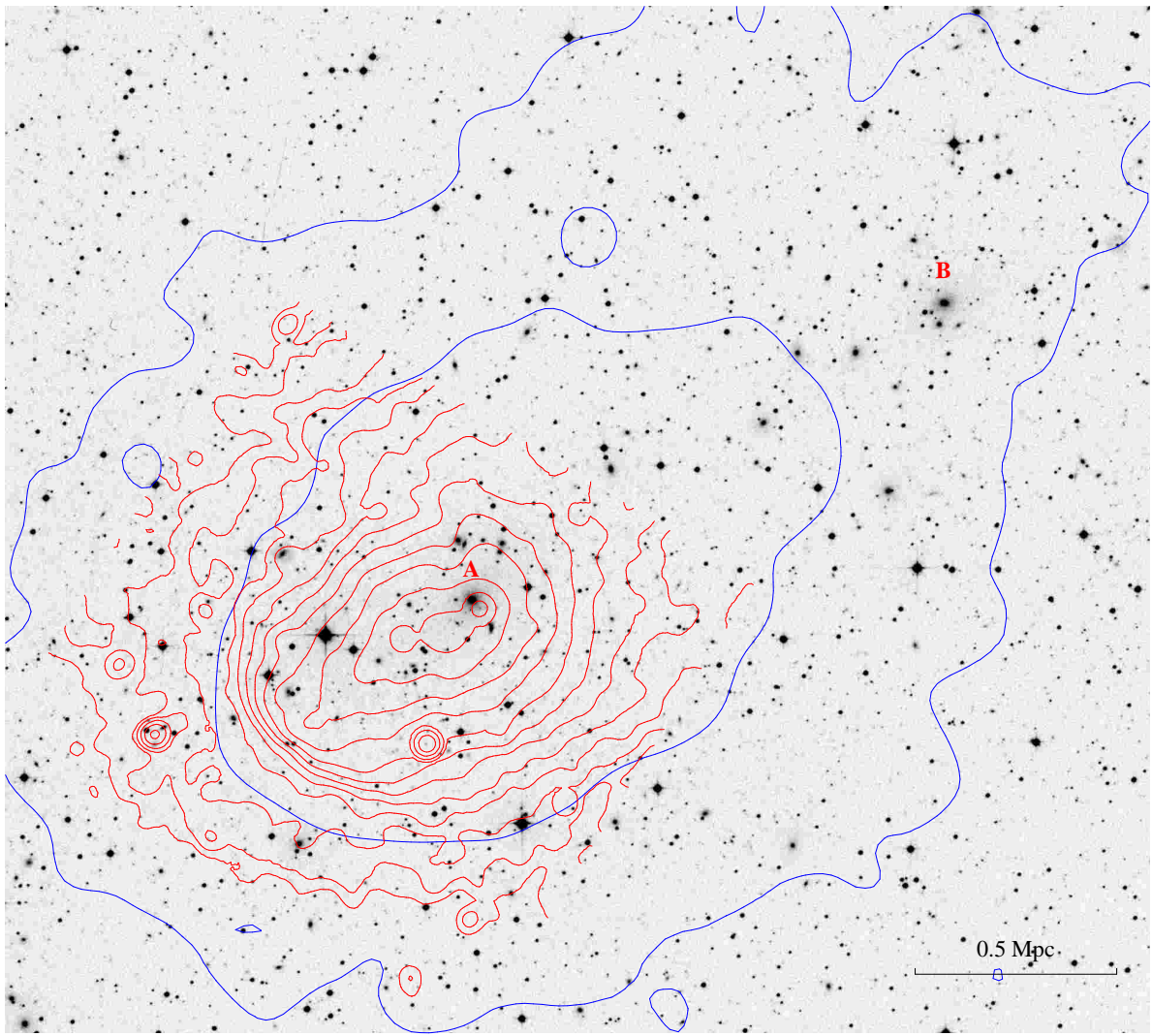}
    }
  \caption{\emph{Chandra} and \emph{ROSAT} X-ray contours overlayed on the
    optical Digitized Sky Survey-II image. Note that the bright point source
    near galaxy A is associated with a faint point-like optical source and
    not with the galaxy. North is up and East is to the left, as they are in
    all other images.
    \label{fig:dss}}
\end{figure*}

Abell 3667 manifests itself in the optical, X-rays, and in the radio as a
spectacular ongoing merger. The cluster optical morphology is bimodal, with
prominent galaxy concentrations around the two brightest galaxies (Sodre et
al.\ 1992). It is likely that these two concentrations (called subclusters A
and B below, see Fig.~\ref{fig:dss}) correspond to the colliding
subclusters. The line-of-sight velocity difference of the dominant galaxies
is only 120~km~s$^{-1}$ (Katgert et al.\ 1998). Since the merger velocities
are expected to be in the range of a few thousand~km~s$^{-1}$, the small
velocity difference between the dominant galaxies strongly suggests that the
merger axis is nearly perpendicular to the line of sight. Bimodal cluster
structure also is evident in the weak lensing mass map (Joffre et al.\ 
2000).  The \emph{ROSAT} PSPC image (Knopp, Henry \& Briel 1996, blue
contours on Fig.~\ref{fig:dss}) suggests an ongoing merger. The diffuse
X-ray emission peaks on the main subcluster and has a general extension
towards the smaller subcluster. In the cluster outskirts, the merger
probably gives rise to the acceleration of relativistic electrons
responsible for the Mpc-scale radio halo (R\"ottgering et al.\ 1997).

The most striking feature in the X-ray surface brightness distribution,
apparent already in the \emph{ROSAT} PSPC and HRI images, is a sharp
arc-like discontinuity in the X-ray surface brightness distribution to the
South-East of subcluster A and oriented approximately perpendicular to the
line connecting subclusters A and B (Markevitch, Sarazin \& Vikhlinin 1999).
The sharpness of this discontinuity led Markevitch et al.\ to suggest that
it was a shock front, even though the coarse \emph{ASCA} temperature map was
not entirely consistent with the shock explanation.

The study of this surface brightness edge was the main goal of our
\emph{Chandra} observation described here. This observation leads to the
unexpected conclusion that the brightness discontinuity is not a shock
front, but the leading edge of a large body of the cool dense gas moving
through the hotter gas, very much like the two structures found with
\emph{Chandra} in A2142 (Markevitch et al.\ 2000).  To reflect the nature of
the surface brightness edge, we will call it a cold front. The high quality
of the \emph{Chandra} data allows a detailed investigation of the cold front
and its vicinity. We show that the dense gas body moves in the hot ambient
gas at nearly the velocity of sound of that gas
(\S~\ref{sec:cloud:velocity}).  We also observe a compression of the hot gas
in front of the cloud (\S~\ref{sec:Stagnation:region}), and further ahead, a
possible bow shock (\S~\ref{sec:bow:shock}). Both these features are
expected to be present in front of a high-velocity cloud. We show that the
front width is smaller than the particle Coulomb mean free path
(\S~\ref{sec:front:width}), and speculate that the diffusion across the
front should be suppressed by formation of a magnetic layer. In the
accompanying paper (Vikhlinin et al.\ 2000), we use the apparent dynamical
stability of the front to derive the strength of the magnetic field in this
layer.

We use $H_0=50$~km~s$^{-1}$~Mpc$^{-1}$ and $q_0=0.5$, which corresponds to
the linear scale of 1.46~kpc/arcsec at the cluster redshift $z=0.055$.

\section{Data Reduction}

A3667 was observed by \emph{Chandra} with the ACIS-I detector in a single
49~ksec observation on Sept 22, 1999, soon after the degradation of the
front-illuminated CCDs has stopped. The data were telemetered in Very Faint
mode. The average total count rate during the quiescent background
intervals, 67.1~cnt~s$^{-1}$, was only slightly below the telemetry
saturation limit of $\sim 69$~cnt~s$^{-1}$ in this mode. This resulted in
data drop-outs during periods of high background. Fortunately, these periods
were isolated in time and affected only a small fraction of the exposure. We
excluded the time intervals when the telemetry was missing in any of the four
ACIS-I CCDs, 6~ksec in total. The remaining 43~ksec of the observation were
not affected by background flares or any other problem.

The standard screening was applied to the photon list, omitting \textsl{ASCA}
grades 1, 5, and 7, known hot pixels, bad columns, and chip node boundaries.
The clean photon list was reprocessed using the latest available calibration
to correct for position-dependent gain variations due to the CTI effect.

The analysis of extended sources requires a careful background subtraction.
Since the particle background on ACIS-I is spatially non-uniform, and also
because A3667 fills the entire field of view, it was necessary to use an
external dataset for background subtraction. The background dataset was
prepared using a procedure described by Markevitch et al.\ (2000) from a
collection of blank field observations with a total exposure of
300--400~ksec, depending on the CCD chip (Markevitch 2000). The accuracy of
the background subtraction was checked using the data at high energies,
where the cluster emission is negligible compared to the background
(10--11~\keV\ in the ACIS-I chips and 5--11~\keV\ in the off-axis ACIS-S2
chip where the cluster surface brightness is much lower). We found that the
background normalization should be increased by 7\%, within the suggested
10\% uncertainty and consistent with the possible long-term trend
(Markevitch 2000). Therefore, in the rest of the analysis, we have increased
the background normalization by 7\% at all energies. All the confidence
intervals reported below include the $\pm10\%$ systematic uncertainty we
allow for the background normalization.

The imaging analysis has been performed in the 0.5--4~keV energy band to
maximize the source-background ratio. The background-subtracted images were
corrected for vignetting and exposure variations.

Because of the CTI effect, the quantum efficiency of the ACIS-I CCDs at high
energies strongly depends on the distance from the chip readout. For
example, the quantum efficiency at 6~keV decreases by $\sim20\%$ at readout
distances greater than $1/3$ of the CCD size. Our spectral analysis uses
approximate position-dependent corrections to the quantum efficiency which
were determined using observations of the supernova remnant G21.5--0.9 and
the Coma cluster (Vikhlinin 2000). For spectra extracted in large regions,
the position-dependent spectral response matrices and effective area
functions were weighted with the cluster flux.

To check the validity of the calibration relevant to the spectral analysis,
we compared the values of cluster temperature measured with \emph{Chandra}
and \emph{ASCA}. The integrated \emph{Chandra} spectrum is well fit by a
Raymond-Smith model with temperature $T=7.31\pm0.17$~keV, heavy metal
abundance $a=0.31\pm0.03$, and absorption $N_H =
(4.1\pm0.3)\times10^{20}\,$cm$^{-2}$ (all uncertainties are at the 68\%
confidence level). The derived temperature is in agreement with the
\emph{ASCA} value, $T=7.0\pm0.4$~\keV\ (Markevitch et al.\ 1998), and the
absorption is reasonably consistent with the Galactic value,
$N_H=4.8\times10^{20}\,$cm$^{-2}$.

\begin{inlinefigure}
  \centerline{\includegraphics{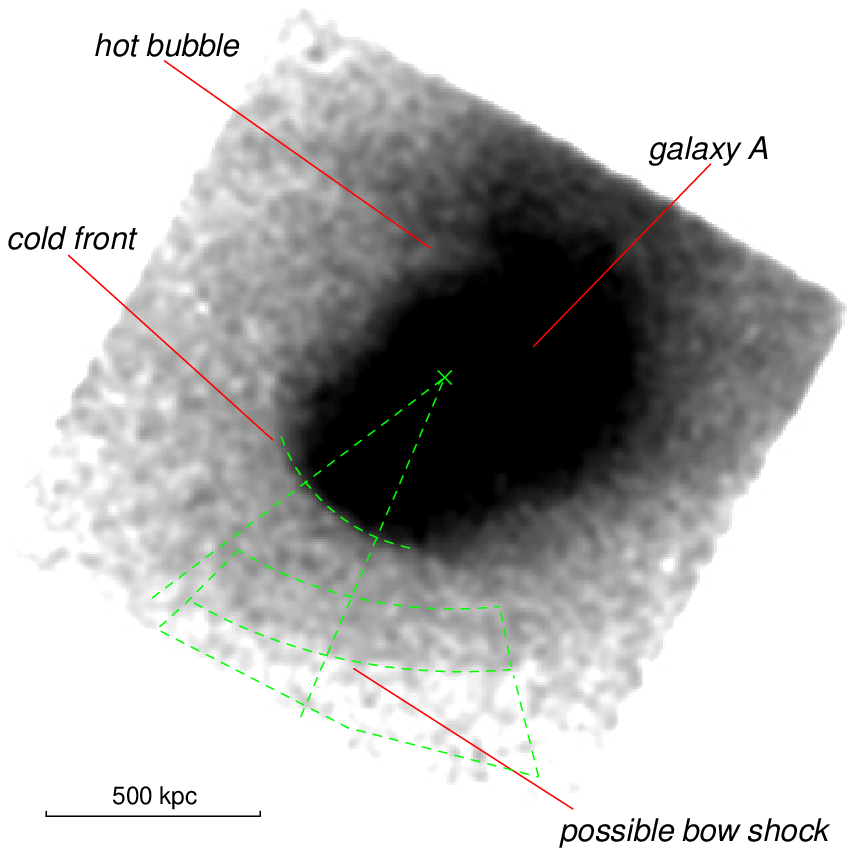}}
  \caption{Smoothed 0.5--4~\keV\ \emph{Chandra} image (full $16'\times16'$
    ACIS-I field of view; North is up, East is to the left). The most
    prominent feature is the sharp surface brightness edge (cold front).
    The front shape is nearly circular as indicated by the arc. The sector
    to the South-East of the center of curvature of the edge (its location
    denoted with an $\times$) was used for the surface brightness and
    temperature profile measurements. The two segments on both sides of the
    possible bow shock indicate the regions where the temperature was
    measured (\S~\ref{sec:bow:shock}).
    \label{fig:img:smo}} 
\end{inlinefigure}

\begin{figure*}[htb]
  \centerline{
    \includegraphics{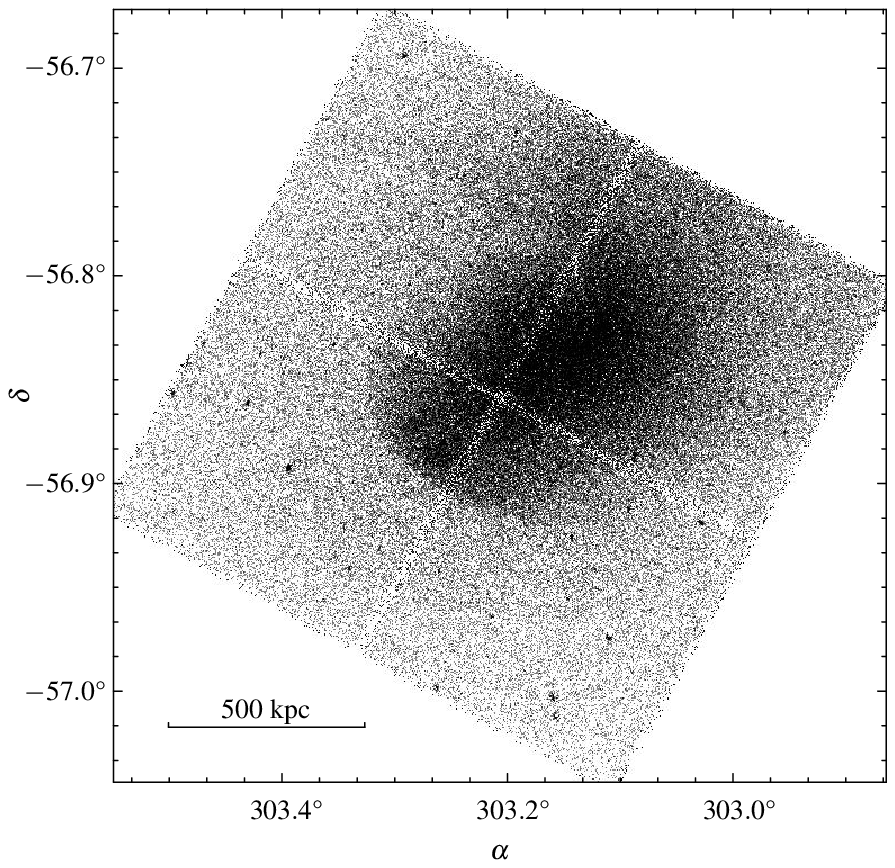}
    \hfill
    \includegraphics{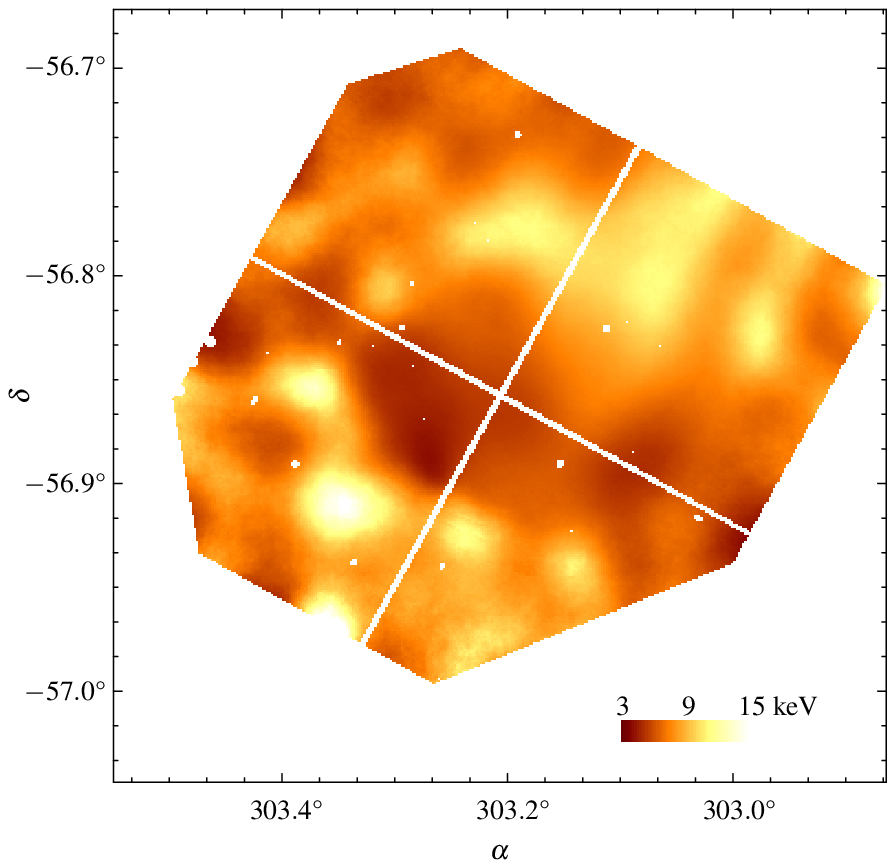}
    }
  \vskip -10pt
  \caption{\emph{(a)} Photon image in the 0.5--4~keV band binned to
    $2''$ pixels. \emph{(b)} Temperature map. The corners are not shown due
    to poor statistics. The typical statistical error in this image is
    $\pm1$~\keV. The cold, $\sim 4$~keV, region near the center of the map
    coincides with the inside of the surface brightness edge. The very high
    temperature in several spots just outside the edge is insignificant. All
    other temperature variations greater than $\pm1$~\keV\ are
    significant.}\label{fig:tmap}
\end{figure*}

\subsection{Temperature Map Determination}

The \emph{Chandra} angular resolution, in principle, allows a
straightforward spectral modeling in very small (a few arcsec) regions,
limited only by statistics.  Our technique for temperature map determination
closely follows that of Markevitch et al.\ (2000). We extracted images in 11
energy bands, 0.7--1.15--1.6--2.5--3--3.5--4.25--5--6--7--8--9~keV bands,
subtracted the background, masked point sources, and smoothed the images
with a variable-width Gaussian filter. The Gaussian filter width was
$\sigma=48''$ everywhere, except for a narrow region near the surface
brightness edge, where the filter width was smoothly reduced to
$\sigma=24''$. We also appropriately weighted the statistical uncertainties
to determine the noise in the smoothed images. The effective area as a
function of energy at each point was calculated as the product of the
position-dependent CCD quantum efficiency and mirror vignetting. Finally, we
generated the response matrices from the position-dependent calibration data
and determined the temperature by fitting these 11 energy points to a
single-temperature plasma model with absorption fixed at the Galactic value
and the heavy element abundance fixed at the best-fit cluster average.

\section{Overview of the X-ray Image and Temperature Map}

The \emph{Chandra} image reveals interesting structures in the central part
of A3667. The features discussed below are indicated in the smoothed image
(Fig.~\ref{fig:img:smo}). The most prominent feature in this image --- and
the main subject of our further investigation --- is the sharp edge in the
surface brightness distribution to the South-East of galaxy~A.  The edge is
even more impressive in the raw photon image (Fig.~\ref{fig:tmap}).  The
edge is located approximately 550~kpc to the South-East of galaxy A, and it
is almost perpendicular to the line connecting subclusters A and B
(Fig.~\ref{fig:dss}).  Its morphology suggests motion to the South-East,
away from galaxy A.  Our temperature map shows that the gas is cooler on the
brighter (denser) side of the edge, which rules out a shock front
interpretation. A detailed analysis of the temperature and density profiles
across this ``cold front'' presented below (\S~\ref{sec:Structure-edge})
shows unambiguously that this is a cold gas body moving to the South-East at
a slightly supersonic speed.

Farther South-South-East of galaxy A, approximately 300~kpc beyond the cold
front, there is a weaker discontinuity in the surface brightness
distribution (Fig.~\ref{fig:img:smo}). We argue in \S~\ref{sec:bow:shock}
that this discontinuity may correspond to a bow shock in front of the
supersonically moving cold gas cloud.

\begin{figure*}[htb]
  \centerline{\includegraphics{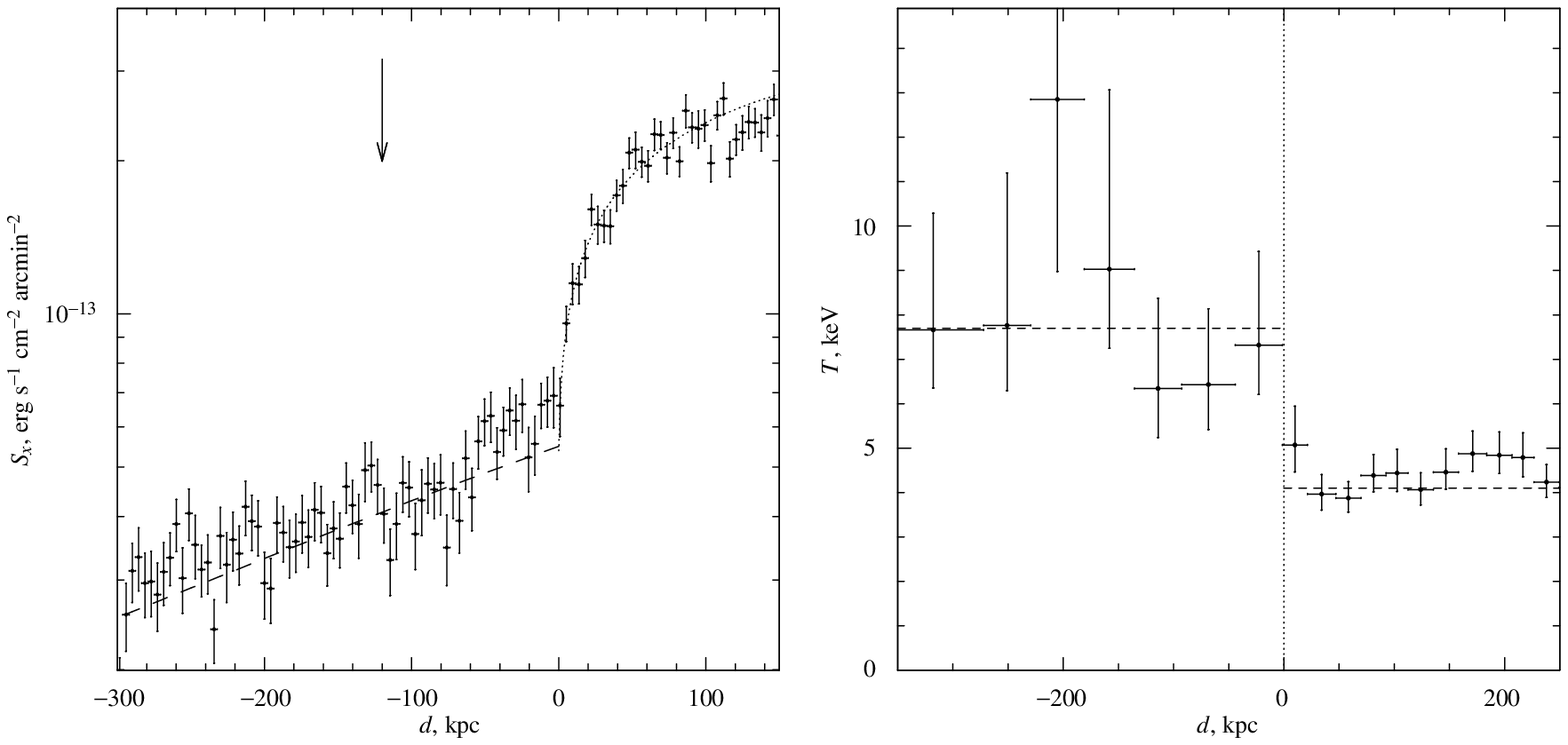}}
  \vskip -10pt
  \caption{ \emph{(a)}: X-ray surface brightness profile (expressed in the
    units of energy flux in the 0.5--2~\keV\ band) across the cold front.
    The profile was measured in the sector shown in Fig.~\ref{fig:img:smo}.
    The \emph{Chandra} photon flux in the 0.5--4~\keV\ band was converted to
    an energy flux in the 0.5--2~keV band. The distance is measured relative
    to the front position. The dashed line shows the \emph{ROSAT} PSPC
    $\beta$-model fit (no renormalization applied) to the profile in the
    outer part of the cluster. The dotted line shows the fit to the
    projected spheroidal density discontinuity model (see text). \emph{(b)}:
    Temperature profile across the front measured in the same sector as the
    surface brightness profile. The dashed lines show the adopted mean gas
    temperature inside and outside the front (see text).
    \label{fig:profiles}}
\end{figure*}

Two additional structures in the temperature map are noteworthy, although
they are not the focus of this paper.  The temperature map reveals a hot
(11~keV) region immediately North-West of galaxy A. This region
coincides with the X-ray surface brightness extension connecting subclusters
A and B. Most likely, the gas in this region is heated by shocks generated
by the merger.

The hot region has an extension at approximately 300~kpc to the North-East
of galaxy A. The extension is almost perpendicular to the merger axis and
coincides with a region of decreased surface brightness, which indicates low
gas density. Such structures can be generated by shock heating of the gas
between the colliding subclusters. If shock heating is strong enough, the
gas specific entropy in this region can be increased so that the gas
becomes convectively unstable. This will result is formation of hot
underdense gas bubbles rising buoyantly along the gradient of the
gravitational potential, i.e.\ perpendicular to the merger axis, which is
what we may be seeing.

Finally, we note that the \emph{Chandra} temperature map reasonably agrees
in the overlapping region with the coarser \emph{ASCA} temperature map
presented in Markevitch et al.\ (1999).

\section{Structure of the Cold Front}
\label{sec:Structure-edge}

In the position angle interval 200\degree--260\degree\ (the position angles are
measured from West through North hereafter) from the galaxy A, where the
surface brightness discontinuity is most prominent (Fig.~\ref{fig:img:smo}),
the isophote of the edge is very well fit by a circle with a radius
$410\pm15$~kpc and a centroid located $\sim 200$~kpc East-South-East of
galaxy A.  The abrupt increase in the X-ray brightness suggests the presence
of a dense gas body with a sharp boundary. Since the surface brightness edge
is nearly circular and since the merger appears to proceed in the plane of
the sky, we can assume that the 3-dimensional shape of the gas body is
reasonably close to a spheroid with the axis of symmetry in the plane of the
sky. These assumptions allow us to derive the gas density within the body by
deprojecting the X-ray surface brightness profile across the edge.

\label{sec:x-ray:profiles}

The X-ray surface brightness and temperature profiles across the edge are
shown in Fig.~\ref{fig:profiles}. They were measured in the
215\degree--245\degree\ sector shown in Fig.~\ref{fig:img:smo}. As we cross the
edge from the outside, the surface brightness increases sharply by a factor
of 2 within 7--10\,kpc and continues to increase smoothly by another factor
of 2 within 40--80\,kpc.  Such a profile suggests a projected density
discontinuity, and is, indeed, well fit by the model of a projected
spheroid.  The surface brightness profile of a spheroid with a constant gas
density is given by
\begin{equation}\label{eq:proj:sphere:text}
  S(d) = 2^{3/2} \sqrt{R}\,\varepsilon_0\;\sqrt{d} \qquad  \mbox{for $d
    \ll R$},
\end{equation}
where $d$ is the distance from the edge, $\varepsilon_0$ is the volume
emissivity of the gas, and $R$ is the radius of curvature of the front (see
Appendix). We add to this equation a component that corresponds to the
projection of the ambient gas. This component is small and nearly constant,
$\approx4.5\times10^{-13}$\,erg~s$^{-1}\,$cm$^{-2}$~arcmin$^{-2}$.  As
Figure~\ref{fig:profiles} shows, this model provides an excellent fit to the
surface brightness within $100$~kpc interior to the edge. The model can be
generalized by using a power law gas density distribution within the
spheroid, $\rho=\rho_0(x^2/b^2+y^2/a^2)^{-3\beta/2}$. The surface brightness
profile in this case is given by eq.~(\ref{eq:proj:ell:pow:law}).  The fit
of this equation to the observed profile 80~kpc interior to the front gives
$\beta=0.07\pm0.08$ (68\% confidence). Therefore, the gas distribution is
consistent with a step-like jump in the density and with a constant density
inside the front. The normalization of the fit provides the electron density
inside the front $n_e=(3.2\pm0.09)\times10^{-3}\,$cm$^{-3}$, where the
uncertainty is due to statistics and the poorly constrained power-law
density slope. A larger contribution to the density uncertainty is the
unknown elongation of the gas cloud along the line of sight.
Equation~(\ref{eq:proj:ell:pow:law:r2}) shows that if the spheroidal cloud
is elongated by a factor of $1+e$ along the line of sight, the gas density
that is needed to reproduce the observed profile decreases as
$(1+e)^{-1/2}$. An elongation $|e|>0.3$ appears unlikely because it would be
perpendicular to the merger axis and because the front shape in projection
is nearly circular. Therefore, the density uncertainty on the inner side of
the front due to the cloud elongation should be within $\approx
e/2=\pm15\%$; this uncertainty is adopted in the discussion below.

The gas density on the outer side of the front was derived using the
\emph{ROSAT} PSPC image because the \emph{Chandra} field of view is too
small to permit a proper deprojection of the slowly varying surface
brightness in this region. The \emph{ROSAT} isophotes are nearly circular in
the South-Eastern sector of the cluster (Knopp et al.\ 1996;
Fig.~\ref{fig:dss}), which suggests that this sector is not significantly
disturbed by the merger. Therefore, the gas density can be derived by the
standard approach assuming spherical symmetry.  The \emph{ROSAT} surface
brightness profile was extracted in the sector 160\degree--250\degree\ centered on
galaxy A and fit with a $\beta$-model, $I=I_0/(1+r^2/r_c^2)^{3\beta-0.5}$,
in the radial range 0.6--3.5~Mpc (i.e., from just outside the front to the
edge of the \emph{ROSAT} field of view).  The $\beta$-model provides a good
fit to the data; the best fit parameters are $\beta=0.79$, $r_c=690$~kpc,
and $I_0=0.0120$~cnt~s$^{-1}$~arcmin$^{-2}$ in the 0.7--2~\keV\ energy
band.  This fit is in excellent agreement with the \emph{Chandra} surface
brightness at distances greater than 70\,kpc of the edge, with no
renormalization required (Fig.~\ref{fig:profiles}). Just outside the front,
the $\beta$-model fit gives the electron density
$n_e=0.82\times10^{-3}\,$cm$^{-3}$. Although the $\beta$-model parameters
are quite sensitive to the choice of the profile centroid, the density at
the edge position is not. For example, if the profile is centered at a point
in the middle of the front, we obtain $\beta=0.73$, $r_c=600$~kpc,
$I_0=2.91\times10^{-3}\,$cnt~s$^{-1}$~arcmin$^{-2}$, and the electron
density near the front is $n_e=0.80\times10^{-3}\,$cm$^{-3}$.

We note that the surface brightness within $\sim70$~kpc exterior to the
front exceeds both the \emph{ROSAT} model and the extrapolation of the
\emph{Chandra} profile from larger radii. Our interpretation of the observed
density edge as a boundary of a fast-moving dense gas body predicts just
such a compression in the immediate vicinity of the front. This is discussed
in more detail in \S~\ref{sec:Stagnation:region}. Our immediate purpose ---
calculation of the cloud velocity --- requires a measurement of the
undisturbed (free stream) gas density on the outer side of the front.  One
approach to derive the free stream density is to use the density just
outside the compression region, at $d\approx-100$~kpc. The \emph{ROSAT}
$\beta$-model gives $n_e=0.71\times10^{-3}\,$cm$^{-3}$ at this distance.  A
better approach is to extrapolate the $\beta$-model to the front position
because it approximately accounts for the small change in the global cluster
gravitational potential. Our discussion below uses the value
$n_e=0.82\times10^{-3}\,$cm$^{-3}$ obtained by extrapolating the
\emph{ROSAT} model to the front position. We allow a $15\%$ uncertainty in
this value.
 
As we cross the edge from the outside, the temperature changes abruptly from
approximately 8\,\keV\ to 4--5\,\keV\ (Fig.~\ref{fig:profiles}). On both
sides, the temperature is consistent with a constant value within some
distance of the front. The fit to the spectra integrated within 275~kpc
beyond and 125\,kpc behind the front are $T_{\rm out}=7.7\pm0.8$~\keV\ and
$T_{\rm in}=4.1\pm0.2$~\keV, respectively. We will adopt these values as
the gas temperature outside and inside the dense cloud.

The temperature in a 10~kpc strip just inside the front, $5.1\pm0.7$~\keV,
is marginally higher than the average inner temperature. Because of the
curvature of the front along the line of sight, this annulus is
significantly affected by projection, with 30--50\% of the emission being
due to the ambient 8\,\keV\ gas. The two-temperature fit in this region,
with the temperature of the hotter component fixed at 8\,\keV\ and a
normalization of 50\% that of the colder, component gives a temperature for
the colder component of $T=3.7\pm0.7$~\keV, which is consistent with the
value at larger distances from the edge. Therefore, the marginal increase of
temperature just inside the front is consistent with projection and does not
require, for example, heat conduction.

In summary, both the density and temperature of the gas change abruptly as
we cross the boundary of the dense cloud.

\begin{inlinetable}
  \def\arraystretch{1.2}
  \caption{Gas parameters in several regions}\label{tab:parameters}
  \begin{tabular}{cccc}
    \hline
    \hline
    Region  & $T$ & $n_e$ & $p=Tn_e$\\
            & \keV & $10^{-3}\,$cm$^{-3}$ & $10^{-2}\,$\keV~cm$^{-3}$\\
    \hline
    Outside the front & $7.7\pm0.8$ & $0.82\pm0.12$ & $0.63\pm0.11$ \\
    Inside  the front& $4.1\pm0.2$ & $3.2\pm0.5$   & $1.32\pm0.21$ \\
    Inside bow shock & $8.2\pm0.9$ & \nodata & \nodata \\
    Outside bow shock & $7.7\pm1.2$ & \nodata & \nodata \\
    \hline
  \end{tabular}
\end{inlinetable}

\section{Discussion}

Let us consider the pressures on the two sides of the cold front. If the
dense gas cloud were at rest relative to the hot gas, the pressures would be
equal on both sides. However, we observe a jump by a factor of $2.09\pm0.48$
(Table \ref{tab:parameters}).  The most natural explanation would be that
the cloud moves through the ambient gas and is subject to its ram pressure
in addition to thermal pressure. From our data, we can determine the
velocity required to produce the additional pressure. Because of the simple
geometry, the gas flow in our case can be approximated by the flow of
uniform gas about a blunt body of revolution. Figure~\ref{fig:sketch} shows
several characteristic regions in such a flow.  Far upstream from the body,
the gas is undisturbed and flows freely. This region is referred to as the
free stream; in the discussion below, we use an index 1 for gas parameters
in this region. Near the leading edge of the body, the gas decelerates
approaching zero velocity at the leading edge.  This region is referred to
as the stagnation point; we will use an index 0 for gas parameters in this
region.  If the velocity of the body exceeds the speed of sound, a bow shock
forms at some distance upstream from the body. The gas parameters just
inside the bow shock will be denoted with an index 2.  Finally, the gas
parameters inside the cloud will have index $0'$.

All these regions are seen in the \emph{Chandra} observation of A3667. In
\S~\ref{sec:cloud:velocity}, we determine the velocity of the dense cloud
from the ratio of gas pressures in the free stream and inside the front. The
cloud velocity turns out to be very close to the velocity of sound in the
ambient gas. In \S~\ref{sec:bow:shock}, we discuss the possible bow shock.
The gas density in front of the fast moving cloud should be significantly
higher than in the free stream. We show in \S~\ref{sec:Stagnation:region}
that this compression is indeed seen in the \emph{Chandra} image.

Most of the further discussion assumes that the gas flow outside the cloud
is adiabatic, i.e., that heat conduction can be neglected. We generally
consider length scales much greater than the Coulomb mean free path; heat
conduction is slower than the gas motions on such scales. Also, and perhaps
more important, heat conduction is likely strongly suppressed by the
presence of magnetic fields.

\begin{inlinefigure}
  \medskip
  \centerline{\includegraphics{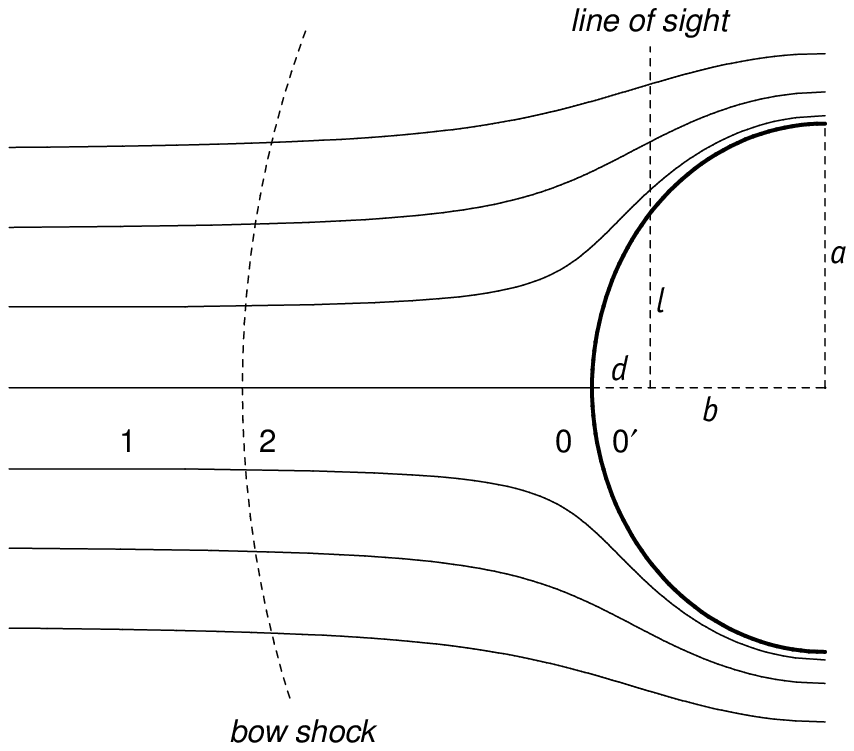}}
  \caption{Geometry of flow past a spheroid with semi-axes
    $a$ and $b$. Zones 0, 1, and 2 are those near the stagnation point, in
    the undisturbed free stream, and past the possible bow shock,
    respectively. Zone $0^\prime$ is within the body.}\label{fig:sketch}
\end{inlinefigure}

\begin{inlinefigure}
  \medskip
  \centerline{\includegraphics{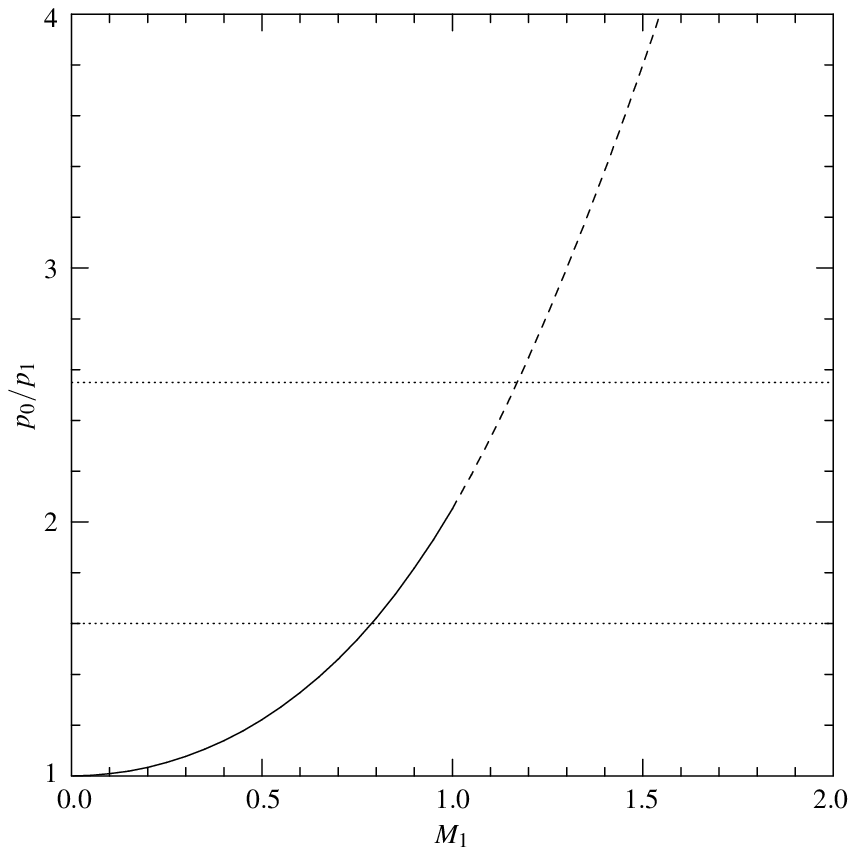}}
  \caption{Ratio of pressures at the stagnation point and in the free stream
    as a function of the Mach number in the free stream. The solid and
    dashed line corresponds to the sub- and supersonic regimes
    (eqs.~\ref{eq:p:ratio:subsonic}, \ref{eq:p:ratio:supersonic}),
    respectively. The dotted lines show the conservative confidence interval
    for the measured pressure ratio.\label{fig:pratio}}
\end{inlinefigure}

\subsection{Velocity of the Hot Gas Cloud}\label{sec:cloud:velocity}

The ratio of pressures in the free stream and at the stagnation point is a
function of the cloud speed $\v$ (Landau \& Lifshitz 1959, \S~114):
\begin{equation}\label{eq:p:ratio:subsonic}
   \frac{p_0}{p_1}=\left(1+\frac{\gamma-1}{2}\;M_1^2\right)^{\frac{\gamma}{\gamma-1}},\quad M_1\le1
\end{equation}
\begin{equation}\label{eq:p:ratio:supersonic}
  \frac{p_0}{p_1}=
        \left(\frac{\gamma+1}{2}\right)^{\frac{\gamma+1}{\gamma-1}}
        M_1^2
        \left[\gamma-\frac{\gamma-1}{2M_1^2}\right]^{-\frac{1}{\gamma-1}},
         \quad M_1>1,
\end{equation}
where $M_1=v/c_1$ is the Mach number in the free stream and $\gamma=5/3$ is
the adiabatic index of the monatomic gas. The subsonic
equation~(\ref{eq:p:ratio:subsonic}) follows from Bernoulli's equation.  The
supersonic equation~(\ref{eq:p:ratio:supersonic}) is more complex because it
accounts for the gas entropy jump at the bow shock.  Figure~\ref{fig:pratio}
illustrates that $p_0/p_1$ is a strong function of the cloud velocity;
therefore the cloud velocity can be measured rather accurately even if the
pressure uncertainties are relatively high.

The gas parameters at the stagnation point --- which enter
eq.~(\ref{eq:p:ratio:subsonic})--(\ref{eq:p:ratio:supersonic}) --- cannot be
measured directly because the stagnation region is physically small and
therefore its X-ray emission is almost completely hidden by projection.
However, the gas pressure at the stagnation point must equal the pressure
within the cloud, $p_{0'}$, which is well-determined. The conservative
confidence interval of the pressure ratio $p_{0'}/\!p_1=2.09\pm0.48$
corresponds to $M_1=1.0\pm0.2$, i.e.\ the gas cloud moving at the sound
speed of the hotter gas. The cloud velocity in physical units is $\v=M_1c_1=
M_1\sqrt{\gamma T/m_p\mu}=1430\pm290$~km~s$^{-1}$, where we used the gas
temperature $T=7.7$~\keV\ and the mean molecular weight of the intracluster
plasma $\mu=0.6$.

\subsection{Possible Bow Shock}\label{sec:bow:shock}

If the speed of a blunt body exceeds the speed of sound, a bow shock forms
at some distance upstream. A surface brightness discontinuity resembling a
bow shock is indeed visible in the \emph{Chandra} image at $\sim 350$~kpc
from the cold front (Fig.~\ref{fig:img:smo}). The shape of this structure is
consistent with an ellipse centered on the center of curvature of the cold
front. The surface brightness profile extracted in concentric elliptical
annuli in the position angle range 230\degree--280\degree\ 
(Fig.~\ref{fig:bowshock}) shows an increase in the surface brightness by a
factor of 1.25 within a narrow radial range around $\sim 350$~kpc from the
cold front. This corresponds to a gas density jump by a factor of $1.1-1.2$,
depending on the projection geometry. If this discontinuity is interpreted
as a shock front, it is straightforward to derive the expected temperature
jump, the shock propagation velocity, and the velocity of the gas behind the
shock, using the Rankine-Hugoniot shock adiabat (Landau \& Lifshitz, \S~85).
For this weak shock with $\rho_2/\rho_1=1.1-1.2$, the propagation velocity
should be close to the velocity of sound in the pre-shock gas, or more
exactly, $1.07-1.13$ of that velocity. The temperature ratio expected for
the shock with such a density jump is $T_2/T_1=1.07-1.13$. This is
consistent with the measured values of $T_2=8.2\pm0.9$~\keV\ and
$T_1=7.7\pm1.2$~keV derived in regions shown in Fig.~\ref{fig:img:smo}.
Evaluating the pre-shock sound velocity from $T_1=7.7$~\keV, we find the
shock propagation velocity $\v\approx1600$~km~s$^{-1}$. As expected for a
stationary shock, this value agrees with the velocity of the cold cloud
derived from independent considerations.

The observed distance between the shock and the cold front agrees well with
that expected for this velocity.  The distance between the bow shock and a
blunt body can be found using the approximate method of Moeckel (1949; see
also Shapiro 1953). The shape of the cold gas cloud in A3667 can be
approximated by a cylinder with diameter $D\sim 500$~kpc and a spherical
head with radius $R=410$~kpc seen as the front in the projection. For this
geometry, Moeckel's method predicts a bow shock distance of 560--320~kpc for
Mach numbers $M_1=1.05-1.2$. Thus the observed distance of 350~kpc
corresponds to a Mach number in the correct range.

We note that the shock-like feature is observed only in the Southern region
of the image and is not symmetric relative to the presumed direction of the
cloud motion.  Since the motion is only slightly supersonic, small-amplitude
temperature nonuniformities can make it subsonic in certain regions and a
shock will not arise everywhere. Another possible explanation is that
nontrivial projection effects mask the shock.

\begin{inlinefigure}
  \medskip
  \centerline{\includegraphics{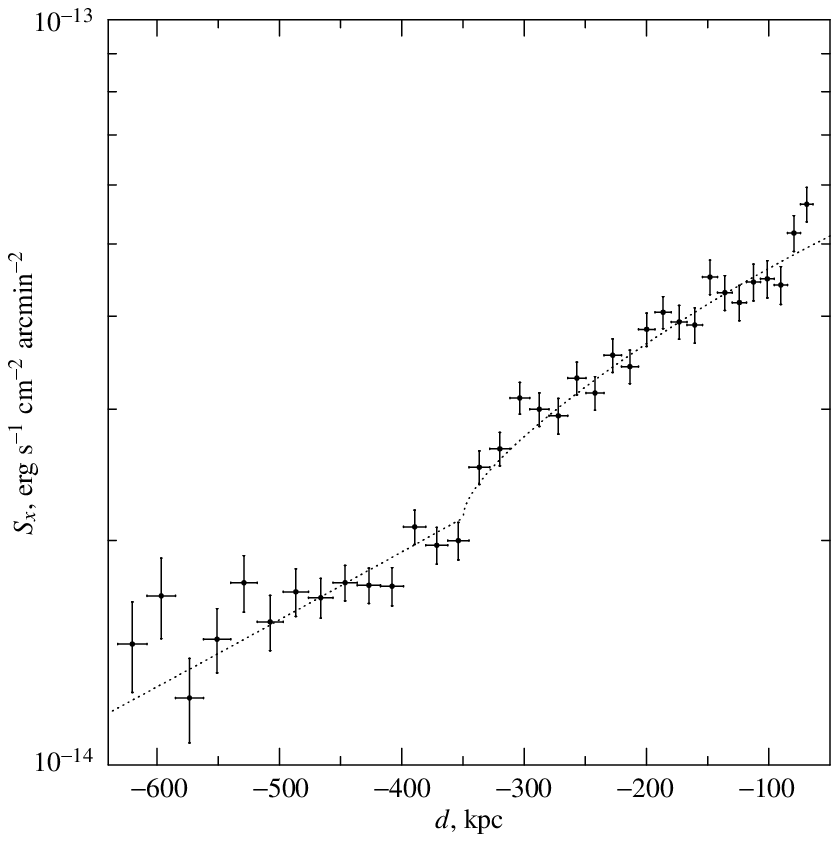}}
  \caption{X-ray surface brightness profile across the possible bow
    shock. The profile has been measured in a sector defined by the two
    segments in Fig.~\ref{fig:img:smo}. The distance is measured from the
    cold front. The dotted line shows a fit to a model in which the gas
    density follows the power law with the same slope on both sides of the
    shock, and undergoes a jump by a factor of 1.15 at the position of the
    shock.
\label{fig:bowshock}}
\end{inlinefigure}

\subsection{Stagnation Region}\label{sec:Stagnation:region}

We now consider the region immediately in front of the dense cloud where the
inflowing gas slows and hence its density rises, and show that this results
in a detectable X-ray brightness enhancement. We will calculate the density
enhancement with respect to either the free stream region, if the cloud
motion is subsonic, or to the region just behind the bow shock, if the
motion is supersonic. In both cases, the Mach number in the reference region
is $M_2\le 1$. The ratio of the gas density, $\rho$, at a point where the
local Mach number of the flow is $M$, and the density in the reference
region, $\rho_2$, is found from Bernoulli's equation (e.g.\ Landau \&
Lifshitz, \S~80), and then the enhancement of the X-ray emissivity,
$\varepsilon$, is
\begin{equation}\label{eq:dens:enhance}
 \frac{\varepsilon}{\varepsilon_2}=\frac{\rho^2}{\rho_2^2}=
  \left(\frac{1+(\gamma-1)M^2/2}{1+(\gamma-1)M_2^2/2}\right)^{-2/(\gamma-1)}.
\end{equation}
If $M_2=1$, the emissivity is enhanced by a factor of 2.4 at the stagnation
point. The integration of eq.~(\ref{eq:dens:enhance}) along the line of
sight gives the enhancement in the X-ray surface brightness. To carry out
the integration, we need to know the gas velocity field. As a first
approximation, we use the velocity field from the numerical simulation of an
$M=1$ flow around a sphere by Rizzi (1980). For a sphere with radius
$R=400$~kpc (the radius of curvature of the cold front) and the observed
gas density, we find the surface brightness enhancement $\Delta S =
2.3\times10^{-14}\,$erg~s$^{-1}$~cm$^{-2}$~arcmin$^{-2}$ with a
characteristic width $\sim\!130$\,kpc in the sky plane. Since the dense
cloud is not a sphere, but rather a cylinder with diameter
$D\approx500$~kpc with a spherical head with $R=400$~kpc, the surface
brightness enhancement has a smaller amplitude and width. The size of the
compressed region along the line of sight is $\approx 2R$ in the case of a
flow around a sphere.  Therefore, for the real geometry, the amplitude of
the surface brightness enhancement should be reduced by a factor $\sim
D/2R=0.7$; the projected width of the compressed region is likely reduced by
a similar factor.  Therefore, we expect the surface brightness enhancement
in front of the cloud to have an amplitude
$1.6\times10^{-14}\,$erg~s$^{-1}$~cm$^{-2}$~arcmin$^{-2}$ and width
$\sim90$~kpc. Quite remarkably, we observe a surface brightness enhancement
with similar parameters (Fig.~\ref{fig:profiles}).

The gas temperature in the stagnation region also should be higher than that
in the free stream due to adiabatic compression. However, the temperature
change is too weak to be detectable. The gas temperature in the adiabatic
flow is proportional to $\rho^{\gamma-1}$. Using
equation~(\ref{eq:dens:enhance}) one finds for $M_2=1$ that the temperature
at the stagnation point is 33\% higher than in the free stream.  Since the
stagnation region contributes only $\sim 30\%$ of the total X-ray brightness
on the outer side of the front (Fig.~\ref{fig:profiles}), the projected
temperature should increase by $\sim 10\%$, which is within the statistical
uncertainties of our temperature measurement.

\begin{inlinefigure}
  \medskip
  \centerline{\includegraphics{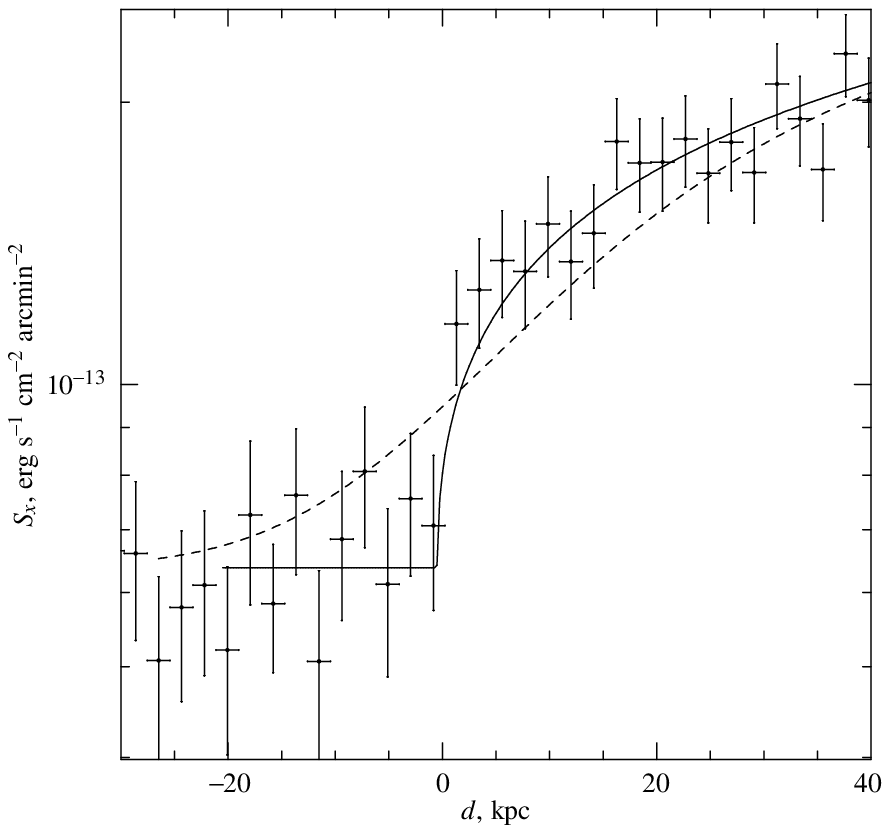}}
  \caption{A detailed X-ray surface brightness profile near the edge
    position. The solid line shows the fit of the projected spherical
    density discontinuity with an infinitely small width. The dashed line
    corresponds to the projection of the same discontinuity smeared with the
    Gaussian $\sigma=20$~kpc.\label{fig:profwidth}}
\end{inlinefigure}

\subsection{Front Width and Suppression of the Transport Processes}
\label{sec:front:width}

The surface brightness edge is remarkably sharp. Indeed, with the
\emph{Chandra} angular resolution, we can tell that its width is smaller
than the Coulomb mean free path of electrons and protons.
Figure~\ref{fig:profwidth} shows a detailed X-ray surface brightness profile
across the front. The surface brightness was measured in narrow
$2.2\times210$~kpc annular segments. The X-ray brightness increases sharply
by a factor of 1.7 within only 2~kpc from the front position. We can compare
this width with the Coulomb mean free path of electrons and protons in the
plasma on both sides of the front. The Coulomb scattering of particles
traveling across the front can be characterized by four different mean free
paths: that of thermal particles in the gas on each side of the front,
$\lambda_{\rm in}$ and $\lambda_{\rm out}$; and that of particles from one
side of the front crossing into the gas on the other side, $\lambda_{\rm
  in\rightarrow out}$ and $\lambda_{\rm out\rightarrow in}$. From Spitzer
(1962) we have for $\lambda_{\rm in}$ or $\lambda_{\rm out}$:
\begin{equation}
  \label{eq:lambda:thermal}
  \lambda = 15~{\rm kpc}\left(\frac{T}{7~{\rm \keV}}\right)^2 
  \left(\frac{n_e}{10^{-3}~{\rm cm}^{-3}}\right)^{-1},
\end{equation}
and for $\lambda_{\rm in\rightarrow out}$ and $\lambda_{\rm out\rightarrow
  in}$:
\begin{equation}
  \label{eq:lambda:in:out}
  \lambda_{\rm in\rightarrow out} = \lambda_{\rm out}\,\frac{T_{\rm
      in}}{T_{\rm out}}\,\frac{G(1)}{G(\sqrt{T_{\rm in}/T_{\rm out}})}
\end{equation}
\begin{equation}
  \label{eq:lambda:out:in}
  \lambda_{\rm out\rightarrow in} = \lambda_{\rm in}\,\frac{T_{\rm
      out}}{T_{\rm in}}\,\frac{G(1)}{G(\sqrt{T_{\rm out}/T_{\rm in}})},
\end{equation}
where $G(x)=(\Phi(x)-x\Phi'(x))/2x^2$ and $\Phi(x)$ is the error function.
Using the gas parameters on the two sides of the front
(Table~\ref{tab:parameters}), we find the numerical values $\lambda_{\rm
  out}=22$~kpc, $\lambda_{\rm in}=1.6$~kpc, $\lambda_{\rm in\rightarrow
  out}=13$~kpc, $\lambda_{\rm out\rightarrow in}=3.5$~kpc. 

The hotter gas in the stagnation region has a low velocity relative to the
cold front. Therefore, diffusion, which is undisturbed by the gas motions,
should smear any density discontinuity by at least several mean free paths
on a very short time scale. Diffusion in our case is mostly from the inside
of the front to the outside, because the particle flux through the unit area
is proportional to $nT^{1/2}$. Thus, if Coulomb diffusion is not suppressed,
the front width should be at least several times $\lambda_{\rm in\rightarrow
  out}$.  Such a smearing is clearly inconsistent with the sharp rise in the
X-ray brightness interior to the front (Fig.~\ref{fig:profwidth}). An upper
limit on the front width can be determined by fitting the observed surface
brightness profile with a model of the projected spherical density
discontinuity whose boundary (in 3D) is smeared with a Gaussian
$\exp(-r^2/2\sigma^2)$. The best fit model (solid line in
Fig.~\ref{fig:profwidth}) requires $\sigma=0$. The formal 95\% upper limit
on $\sigma$ is only 5~kpc. This can be explained only if the diffusion
coefficient is suppressed by at least a factor of 3--4 with respect to the
Coulomb value.

The suppression of transport processes in the intergalactic medium is
usually explained by the presence of a magnetic field. The suppression is
effective if the magnetic field lines are either strongly tangled or there
is a large-scale field perpendicular to the direction in which diffusion or
heat conduction is to occur. If the magnetic field is tangled, the observed
width of the front, $\sigma<5$~kpc, is the upper limit on the magnetic loop
size. The other possibility is that the magnetic field is mostly parallel to
the front at least in a narrow layer. In the subsequent paper (Vikhlinin et
al.\ 2000), we show that the existence of such a parallel component in the
magnetic field is required for dynamical stability of the front. Such a
component would provide effective isolation of the cold gas cloud and
prevent the observed sharp density gradient from dissipating.

\section{Summary}

The analysis of \emph{Chandra} observation of A3667 presented here leads to
the following conclusions:

1. The prominent sharp discontinuity in the X-ray surface brightness
distribution spanning 0.5~Mpc around the South-Eastern side of one of the
two subclusters, is not a shock front as proposed before, but instead a
``cold front'', i.e., a boundary of a dense cool gas body embedded in the
hotter ambient gas.

2. The gas pressure inside the cloud is $2.1\pm0.5$ times higher than that
in the ambient gas. The higher pressure inside the cloud is most plausibly
explained by the ram pressure caused by the motion of the cloud. The
amplitude of the pressure jump requires near-sonic velocity of the cloud
relative to the ambient gas (the Mach number is $M=1\pm0.2$)

3. The near-sonic velocity of the cloud is independently confirmed by
observation of compression of the ambient gas near the leading edge of the
cloud, and by a possible bow shock seen $\sim 350$~kpc ahead of the cold
front.

4. The width of the cold front is smaller than $3.5''$ or 5~kpc, which is a
factor of 2--3 smaller than the Coulomb mean free path. This means that we
directly observe the suppression of transport processes in the intergalactic
medium, most likely by a magnetic field.

\acknowledgments

The results presented here are made possible by the successful effort of the
entire \emph{Chandra} team to build, launch, and operate the observatory. We
acknowledge helpful discussions with E.~Churazov, W.\ Forman, and C. Jones.
This study was supported by NASA grant NAG5-9217 and contract NAS8-39073.

\begin{appendix}

\section{Projection of the Spheroidal Gas Cloud}
\label{sec:Proj-spher-fill}

Projection of an abrupt density jump at the boundary of the spheroidal gas
cloud results in a sharp elliptical edge in the surface brightness
distribution. Here we calculate the expected surface brightness profile
across this edge. The geometry of the problem and the notations are defined
in Fig.~\ref{fig:sketch}. We assume the gas density near the spheroid
boundary follows a power law
\begin{equation}
  \rho(x,y) = \rho_0\left(\frac{x^2}{b^2}+\frac{y^2}{a^2}\right)^{-3\beta/2},
\end{equation}
where $\rho_0$ is the value at the boundary. The volume emissivity
distribution is the square of the density distribution
\begin{equation}
  \varepsilon = \varepsilon_0\left(\frac{x^2}{b^2}+\frac{y^2}{a^2}\right)^{-3\beta}.
\end{equation}
The line-of-sight distance through the spheroid, $l$, at the projected
distance $d$ from the boundary is
\begin{equation}
  l = a \,\left(1-\frac{(b-d)^2}{b^2}\right)^{1/2} =
  a \,\left(\frac{2d}{b}-\frac{d^2}{b^2}\right)^{1/2}.
\end{equation}
The surface brightness profile then is
\begin{equation}\label{eq:proj:ell:pow:law}
  S(d) = \int_{-l}^l \varepsilon\,dy = 
  2a\varepsilon_0\;\left(\frac{2d}{b}-\frac{d^2}{b^2}\right)^{1/2} \;\times
  \;\left(\frac{b-d}{d}\right)^{-6\beta}\;
  \HF\left[\frac{1}{2},3\beta,\frac{3}{2};\frac{d(d-2b)}{(d-b)^2}\right],
\end{equation}
where $\HF$ is the hypergeometric function. The first factor in this equation,
$2a\varepsilon_0\sqrt{2d/b-d^2/b^2}$, is the surface brightness profile
expected for constant gas density, and the rest is the correction due 
to the power-law density distribution. For $|\beta|<0.25$, this correction
factor can be adequately approximated as
$(1-d/b)^{-3.45\beta}$.

We are interested in the surface brightness profile at distances $d$ small
compared to the spheroid axis $b$. In this case, one can neglect the term
$d^2/b^2$ under the square root as well as the correction due to the power
law density profile. Equation~(\ref{eq:proj:ell:pow:law}) then can be
simplified to
\begin{equation}\label{eq:proj:ell:pow:law:r}
  S(d) \approx
  2^{3/2} \sqrt{R}\,\varepsilon_0\;\sqrt{d} \qquad  (d \ll b),
\end{equation}
where $R$ is the radius of curvature of the ellipse near the edge,
$R=a^2/b$. 

Finally, we should take into account a possible deviation of the gas cloud
shape from the spheroid. A reasonable generalization is to assume that the
cloud is ellipsoidal with one of the axes parallel to the line of sight. In
this case, the semi-axis $a$ in eq.~(\ref{eq:proj:ell:pow:law}) cannot be
derived from observations. Instead, the shape of the surface brightness edge
in projection defines $b$ and the third semi-axis, $a'$. If we write
$a=(1+e)a'$, where $e$ is the unknown elongation of the cloud along the line
of sight, eq.~(\ref{eq:proj:ell:pow:law:r}) becomes
\begin{equation}\label{eq:proj:ell:pow:law:r2}
  S(d) \approx
  2^{3/2} (1+e) \sqrt{R_1}\,\varepsilon_0\;\sqrt{d}\qquad  (d \ll b),
\end{equation}
where $R_1$ is the radius of curvature of the brightness edge in projection.
Therefore, a projected ellipsoid should still have the characteristic square
root surface brightness profile near the edge. The normalization of the
profile depends only on the volume emissivity of the gas, the radius of
curvature of the edge in projection, and on the elongation along the line of
sight. Given the observed profile $S(d)$, the gas density is
\begin{equation}
  \rho\approx\rho_0\propto\sqrt{\varepsilon_0}=2^{-3/4}\sqrt{S(d)}\;
  d^{-1/4} \times R_1^{-1/4}(1+e)^{-1/2}.
\end{equation}
It depends only weakly on the geometrical factors $R_1$ and $e$.

\end{appendix}


\begin{references}

  \reference{} Gerwin, R. A. 1968, Rev Mod Phys, 40, 652

  \reference{} Gooderum, P. B. \& Wood, G. P. 1950, NACA Technical Note 2173

  \reference{} Joffre, M. et al.\ 2000, ApJ, 534, L131

  \reference{} Katgert, P., Mazure, A., den Hartog, R., Adami, C., Biviano,
  A., \& Perea, J. 1998, A\&AS, 129, 399
  
  \reference{} Knopp, G.~P., Henry, J.~P., \& Briel, U.~G. 1996, ApJ, 472,
  125

  \reference{} Landau, L.~D., \& Lifshitz, E.\ M.\ 1959, Fluid Mechanics
  (London: Pergamon)

  \reference{} Markevitch, M., Forman, W., Sarazin, C. L., \& Vikhlinin,
  A. 1998, ApJ, 503, 77.

  \reference{} Markevitch, M., Sarazin, C.~L., \& Vikhlinin, A. 1999, ApJ,
  521, 526.

  \reference{} Markevitch, M. 2000, http://asc.harvard.edu
  Calibration$\rightarrow$ACIS$\rightarrow$ACIS background

  \reference{} Moeckel, W. E. 1943, Approximate Method for Predicting Form
  and Location of Detached Shock Waves, NACA Technical Note 1921

  \reference{} Rizzi, A. 1980, in Numerical Methods in Applied Fluid
  Dynamics, ed. B. Hunt (London: Academic Press), p~555

  \reference{} R\"ottgering, H. J. A., Wieringa, M. H., Hunstead, R. W., \&
  Ekerts, R. D. 1997, MNRAS, 290, 577

  \reference{} Shapiro, A. H. 1953,  The Dynamics and Thermodynamics of
  Compressible Fluid Flow (New York: Ronald Press) \S~22.6

  \reference{} Sodre, L., Capelato, H. V., Steiner, J. E., Proust, D., \&
  Mazure, A.\ 1992, MNRAS, 259, 233

  \reference{} Vikhlinin, A. 2000, http://asc.harvard.edu
  Calibration$\rightarrow$ACIS$\rightarrow$CTI-Induced Quantum Efficiency Loss

  \reference{} Vikhlinin, A., Markevitch M., \& Murray, S. S. 2000, ApJ
  Letters, submitted (astro-ph/0008499)

\end{references}
\end{document}